\documentclass{appolb}
\usepackage{epsfig}

\begin{document}
\title{The Heider balance and social distance%
\thanks{Presented at the First Polish Symposium on Econo- and Sociophysics, Warsaw, November 2004.}%
}
\author{P. Gawro{\'n}ski, P. Gronek and K. Ku{\l}akowski
\address{Faculty of Physics and Applied Computer Science,
AGH University of Science and Technology\\
al. Mickiewicza 30, PL-30059 Krak\'ow, Poland}
}
\maketitle
\begin{abstract}

The Heider balance is a state of a group of people with established mutual relations between them. 
These relations, friendly or hostile, can be measured in the Bogardus scale 
of the social distance. In previous works on the Heider balance, these relations have 
been described with integers $0$ and $\pm1$. Recently we have proposed real numbers instead.
Also, differential equations have been used to simulate the time evolution of the 
relations, which were allowed to vary within a given range. In this work, we investigate 
an influence of this allowed range on the system dynamics. As a result, we have found that a 
narrowing of the range of relations leads to a large delay in achieving the Heider balance.
Another point is that a slight shift of the initial distribution of the social distance
towards friendship can lead to a total elimination of hostility.
\end{abstract}
\PACS{ 87.23.Ge }
  
\section{Introduction}

There is a long debate about essential differences between social and natural sciences,
and in particular between sociology and physics \cite{mer,col}. In these discussions, 
the term 'physics' is used with its most deterministic and methodologically clear
part in mind, i.e. the Newton equations. However, the proud claim of Newton: "Hypotheses
non fingo" can hardly be repeated, when talking about any branch of modern physics. In particular,
statistical mechanics is known to be a phenomenological science and the validity 
of its results should always be confirmed by experiment \cite{kubo}. Statistical mechanics 
is maybe a branch which is most comparable to sociology, for the great role of the law 
of large numbers in both these sciences. On the other hand, in sociology there is no 
unique and commonly accepted method, but rather a rich set of schools, grown within
different intellectual traditions \cite{szac}. Its most mathematical part, the game theory, was 
initialized by Janos von Neumann in 40's \cite{neu,str}. However, most of the 
game theory is oriented rationally, whereas the human mind is not \cite{bau}. It is of 
interest to develop a quantitative theory of society, which could include this aspect 
of ours. Such a theory was initialized (again in 40's) by Fritz Heider \cite{hei1,hei2,kad}. 

The Heider balance (HB) is a concept in social psychology \cite{ntc}. For
the purposes of this work, the following description will be sufficient. A group of
people is initially connected with relations which are randomly hostile or friendly.
These random relations cause a cognitive dissonance: which can be of two kinds.
First is when the relations of three people are mutually hostile. Each member of this
triad is willing to decide whom he dislikes less, and in turn to improve relations with him.
As this kind of relations usually tends to be symmetric, the dissonance is removed
by a friendship between two persons, third one being excluded. Second case of the 
dissonance within a triad is when two persons dislike each other, but both like a third
one. A mutual hostility between good friends of someone seems strange: if both are OK,
why they are as stubborn as not to recognize it? Usually either we look for someone guilty
and we like him less, or we try to invite them to a consent. Then again, the stable 
situation in a triad is either two hostile links, or a friendship. On the contrary, 
an unbalanced triad is of one or three hostile relations. This recalls the concept of 
frustration in spin glasses \cite{rev}; indeed, the product of three links is negative 
for an unbalanced triad, positive for a balanced one.

As it was early recognized \cite{har}, a gradual modification of the unbalanced triads 
leads the whole system towards a separation of two subsets, with friendly relations 
within each set and hostile relations between the sets. This is the so-called first 
structure theorem. For more detailed discussion of this stage of work see Ref. \cite{dor}.
Removing of the cognitive dissonance in the sense explained above leads then to a
kind of social mitosis \cite{wang}. Algorithm of repairing the unbalanced triads
was used in \cite{wang} to investigate the role of initial distribution of relations,
described as $\pm 1$ and eventually zero. The approach was generalized recently \cite{my}
by the present authors with using real numbers instead of integers. In this model, each 
relation between group members $i=1,...,N$ is modelled by a matrix element $r(i,j)\in(-R,R)$.
Such a relation is equivalent to a kind of social distance between $i$ and $j$, and it can
be measured with polls in the Bogardus scale \cite{bg}. For brevity, we denote $r(i,j)\equiv r$ 
from now on. 

The range $(-R,R)$ is a limitation of relations from below and from above, and its use is 
motivated by the fact that our feelings affirmed openly are usually moderate, not to insult 
the public opinion. In Ref. \cite{my}, $R=5.0$ was used, and the initial distribution
of $r(i,j)$ was $\rho(r)=1.0$ for $r\in(-0.5,0.5)$. The difference between initial and 
ultimate limit, i.e. between $0.5$ and $5.0$, was found to be large enough not to influence 
the time and dynamics of achieving the Heider balance, which were the same as for $R=\infty $.
However, it is of interest to investigate how the dynamics changes when $R$ decreases.
Tightening of allowed range of the relations $r$ is interpreted here as {\it Gedankenexperiment},
where the public opinion becomes more restrictive. To observe possible consequences of such
a tightening is one of the aims of this work. Another aim is to check an ifluence of the initial 
probability distribution of $r$ on the process dynamics. The same distribution is used for 
all matrix elements $r(i,j)$.

In subsequent section we recapitulate our formulation of the problem of the Heider balance 
\cite{my}, including the equations of motion of the matrix elements $r(i,j)$. Section III contains 
new numerical results, which are discussed in the last section.

\section{The model}

In the simplest picture, the group members can be visualised as nodes of a fully connected graph,
and the relations between them - as links between the nodes. These relations are described by the 
matrix elements $r(i,j)$, $i,j=1,...,N$. The proposed equations of time evolution of the relations
\cite{my} are 

\begin{equation}
\frac{dr(i,j)}{dt} = G[r(i,j),R] \sum_k r(i,k) r(k,j)
\end{equation}
with $G[r,R]=1-(r/R)^2$ as a multiplicative factor which introduces the limitation of
$r(i,j)$. For $r(i,j)$ out of the range $(-R,R)$, the time derivative of $r(i,j)$ is set
to be zero. Initially, the matrix elements $r(i,j)$ are random numbers uniformly distributed
in the range $(-r_0,r_0)$, where $r_0<R$. In this way, the numbers $r(i,j)$ will never leave 
the range $(-R,R)$.
The direct form of the function $G[r,R]$ is of minor importance here, as well as the assumed 
values of $r_0$ and $R$. We guess that the results should scale with $r_0/R$.
For computational reasons, the function $G$ is chosen to be as elementary as possible. 
Diagonal elements $r(i,i)$ are zero. For the sociological interpretation, the summation over 
the nodes $k$ in Eq. 1 is crucial. It means, that the velocity of change of the relations
between $i$ and $j$ is determined by the relations of $i$ and $j$ with all other group members.
If both the relation between $i$ and $k$ and the relation between $k$ and $j$ are friendly or 
both of them are hostile, the relation between $i$ and $j$ is improved. On the contrary, 
if $(i,k)$ are friends and $(k,j)$ are enemies, the relation between $i$ and $j$ gets worse.
The ruling principle is: 'my friend's friend is my friend, my friend's enemy is my enemy, 
my enemy's friend is my enemy, my enemy's enemy is my friend' \cite{hei2,sit}.
Unlike the classic formulation in terms of integers \cite{wang}, here the above velocity depends
also on the intensity of the relations, and not only on their sign.

\section{The results}

In Ref. \cite{my} we have demonstrated that for large systems the time dependence of the
number of unbalanced triads is similar to the theta fuction: initially flat, after some time
abruptly goes to zero. This kind of dynamics of a social system can be considered as unwelcome,
if we remember that the Heider balance in a fully connected graph is accompanied to the division
of the group into mutually hostile camps. Also, as a byproduct of this kind of removing of 
cognitive dissonance we get the relations polarized. This means that members of each group
strongly believe that their group is right, while the other is wrong.
We are interested in an influence of the narrowing of the range $(-R,R)$ on the time of reaching 
the Heider balance and on the character of the process. The interest is motivated by the following
question: to what extent can the group unity be preserved if the dynamics of the relations is
bound? It is obvious that in the trivial limit of $r_0=R=0$, the relations do not evolve at all.
However, as soon as $R>0$, the fixed point $r=0$ is not stable \cite{my}. 

\begin{figure}
\begin{center}
\includegraphics[angle=-90,width=.8\textwidth]{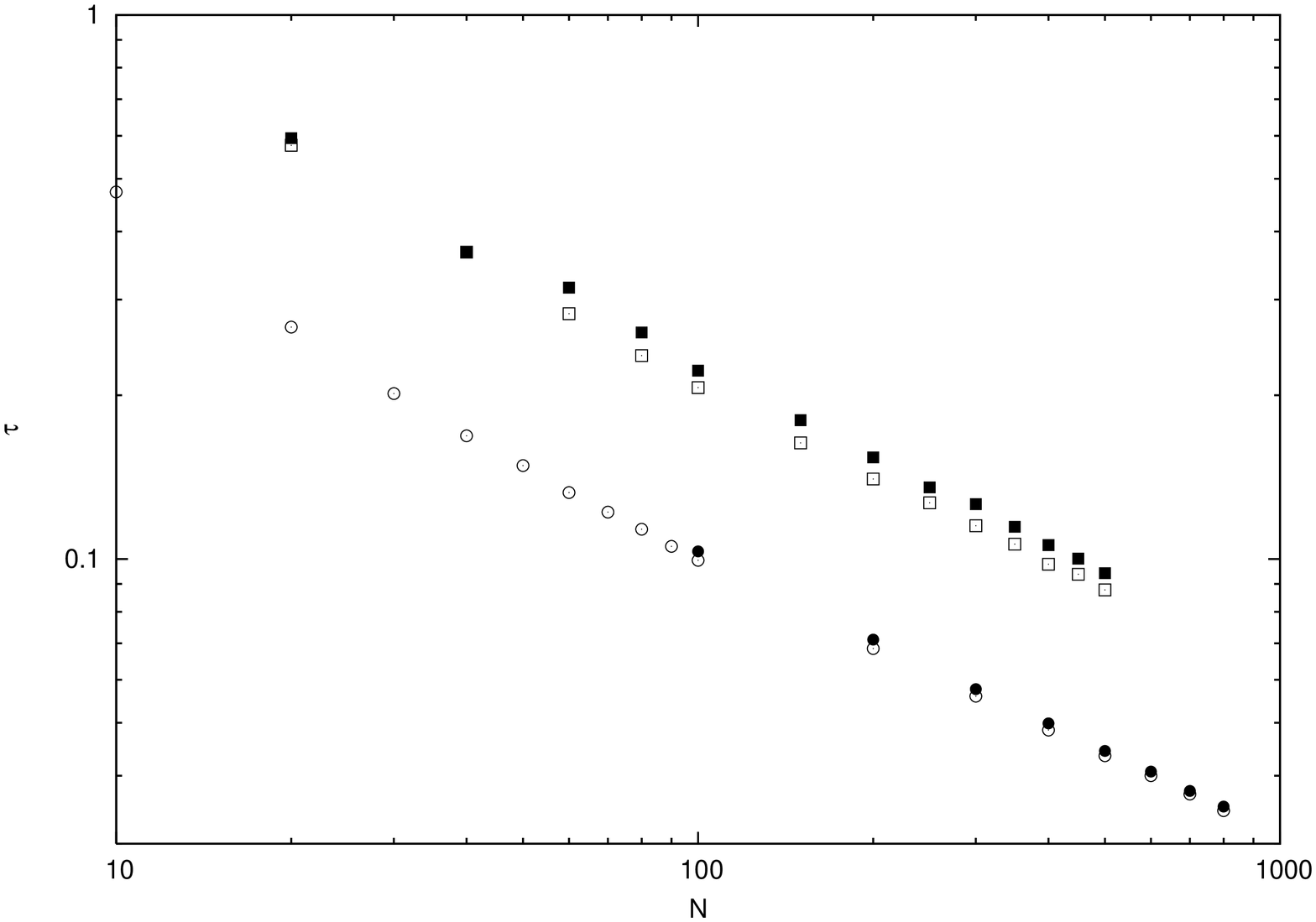}
\includegraphics[angle=-90,width=.8\textwidth]{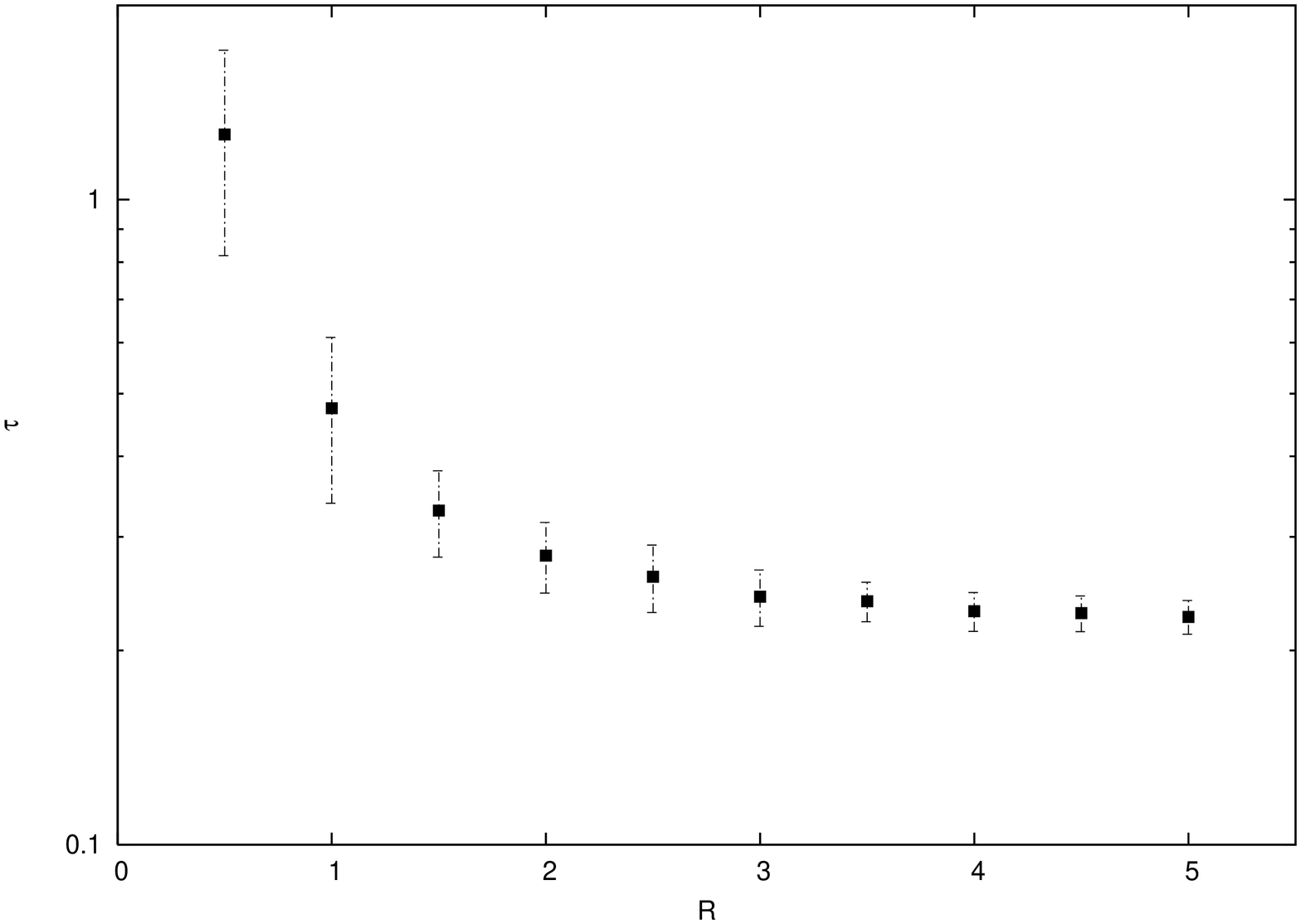}
\caption{Time $\tau$ of reaching the Heider balance as dependent on (a) the system size $N$,
for $R=5.0$ (black symbols) and $R=\infty $ (empty symbols). The higher curve is for $r_0=0.5$, 
the lower is for $r_0=1.0$; (b) the limit value $R$ for $N=100$. Above $R=3$, $\tau$ 
changes very slowly. }
\label{tau_N_R}
\end{center}
\end{figure}

In Fig. 1 a, we show the time $\tau$ of getting the Heider balance as dependent on the system size 
$N$, for various $R$. Here, the results are obtained for systems somewhat larger, than it was 
shown in Ref. \cite{my}. The results point out that $\tau$ decreases with $R$, i.e. the whole 
process takes more time if the relations are more limited. The same rule is demonstrated to be 
true in Fig. 1 b, where we show an example of the dependence $\tau (R)$ for a given system
size $N$. It seems that this time can go to infinity if $R$ is small enough.

\begin{figure}
\begin{center}
\includegraphics[angle=-90,width=.8\textwidth]{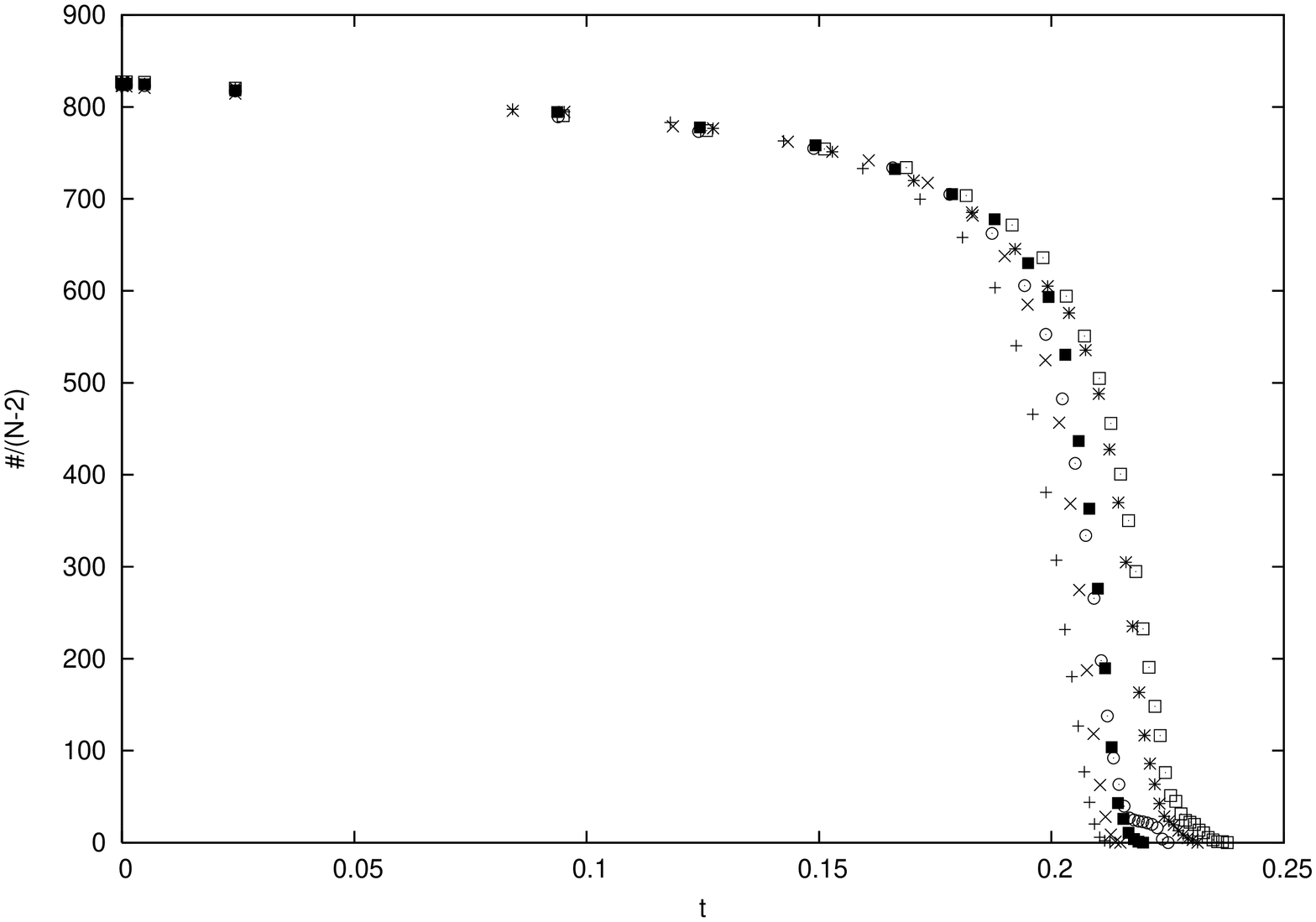}
\includegraphics[angle=-90,width=.8\textwidth]{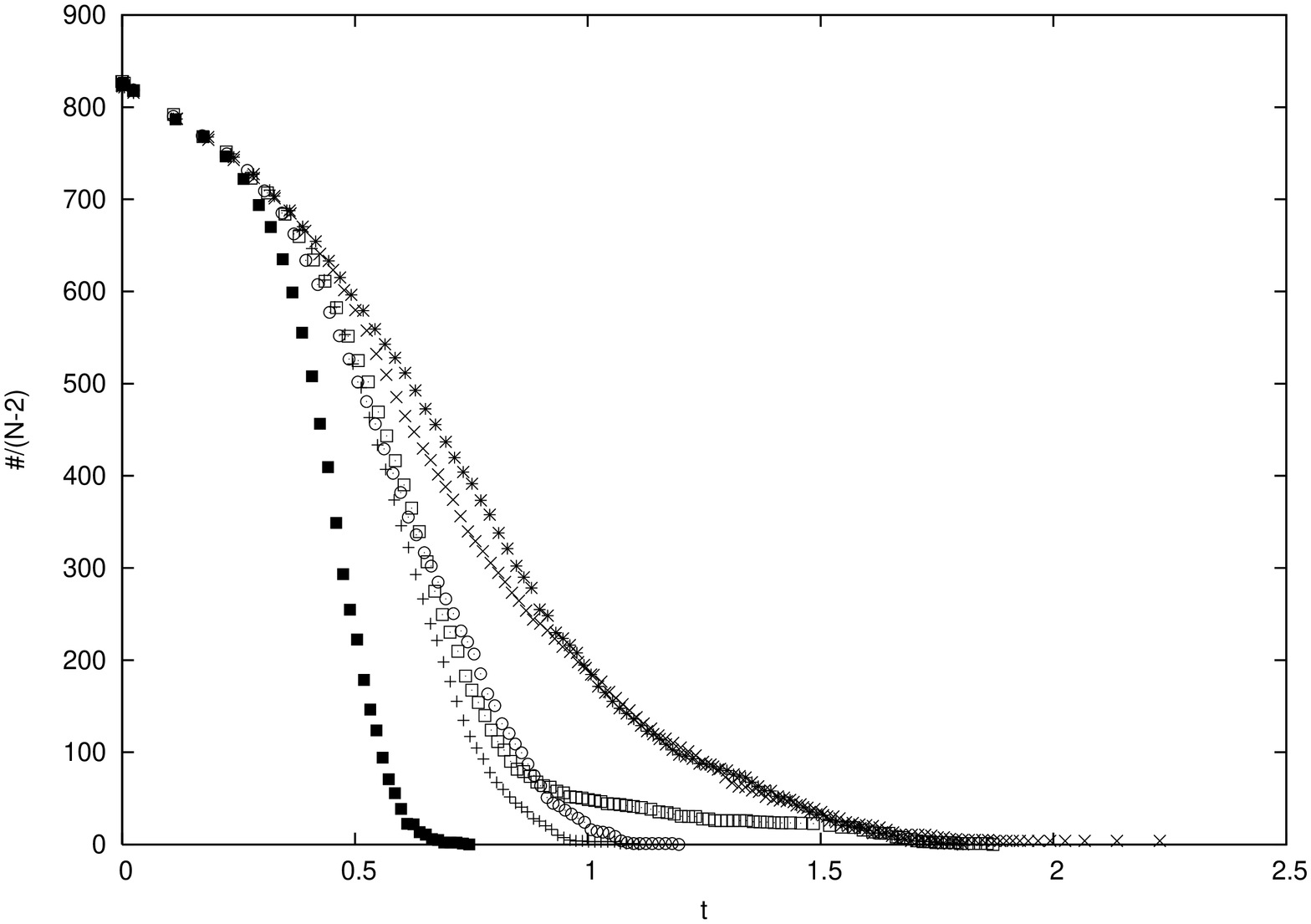}
\caption{Examples of time dependence of the number $\#$ of unbalanced triads for $N=100$ 
and (a) $R=5.0$,
(b) $R=0.5$. The vertical scale of is set as to have $\#/(N-2)=1$ for its minimal possible value,
where the sign of only one link prevents the system to be in the Heider balance. Note different 
scales in (a) and (b).}
\label{NN_t}
\end{center}
\end{figure}

We have also analyzed the system dynamics as dependent on the value of $R$. As it was remarked 
in Ref. \cite{my}, for $N=100$ and higher the time dependence of the number $N_{min}$ of 
unbalanced triads decreases
abruptly just before the balance is achieved. Some examples of such a course is given in Fig. 2 a
for $R=5.0$. In the same conditions but for $R=1.0$, the time $\tau$ is much longer. Moreover,
in the last stage the number $N_{min}$ changes very slowly. Actually, for one of the examples 
shown in Fig. 2 b $\tau=3.5$. In Figs. 2 a, b the vertical axis is in units $\#/(N-2)$, where 
$\#$ is the number of unbalanced triads. This scale is selected as to obtain unity when the sign 
of only one link is different than it should be in the balance state. In this scale we can easily
notice that the number of unbalanced triads varies quickly indeed in Fig. 2 a, but quite 
slowly in Fig. 2 b.

\begin{figure}
\begin{center}
\includegraphics[angle=-90,width=.8\textwidth]{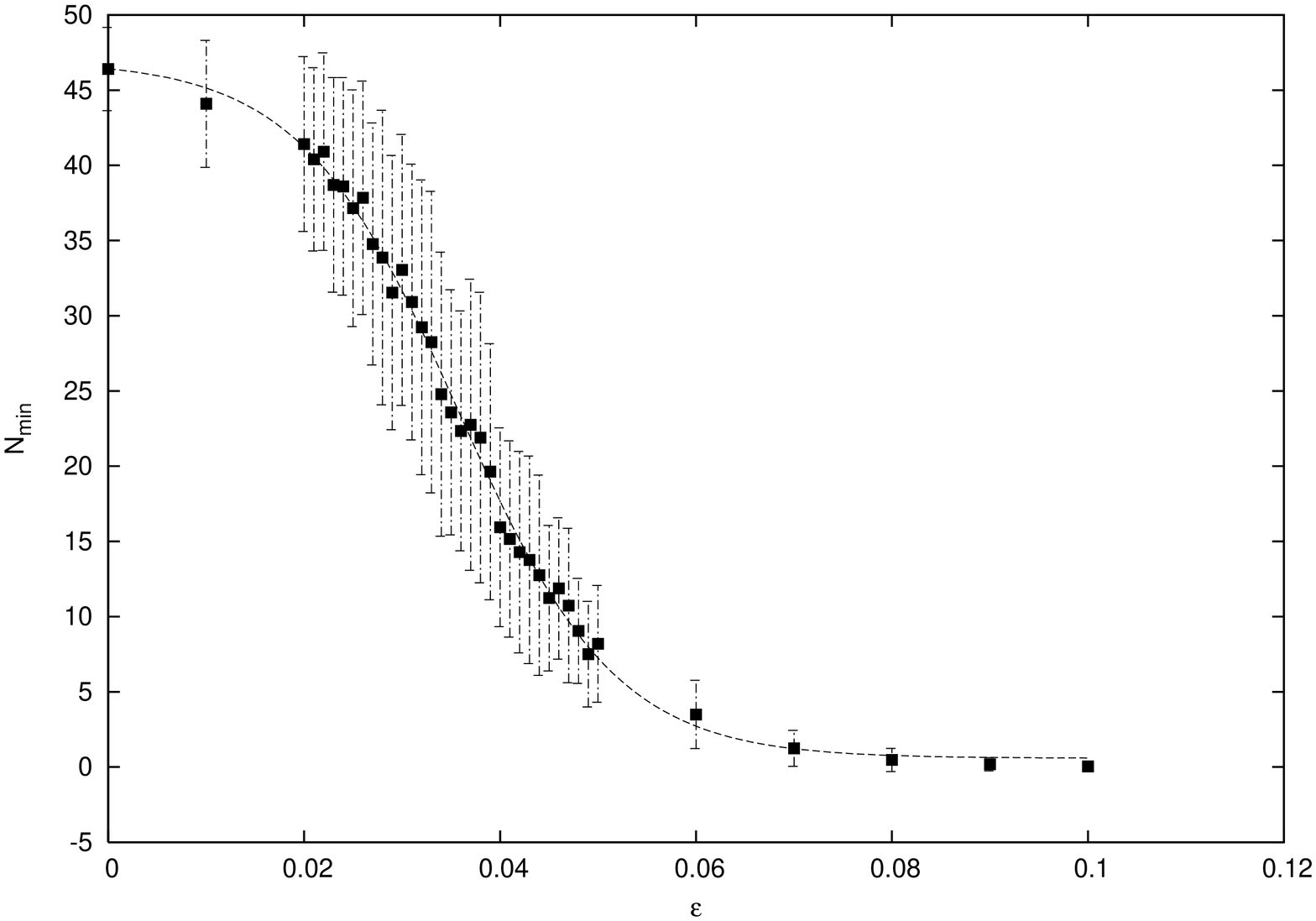}
\includegraphics[angle=-90,width=.8\textwidth]{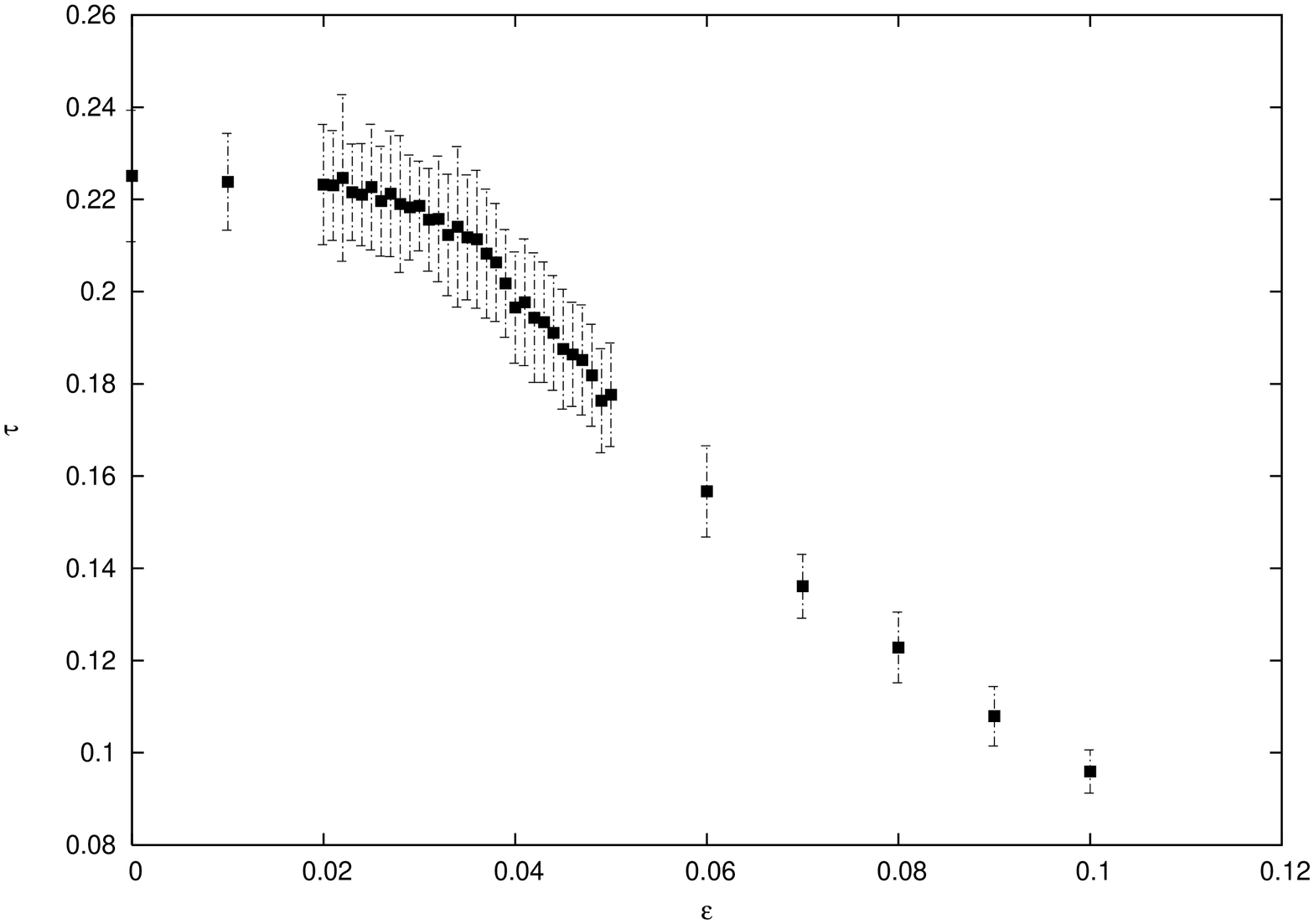}
\caption{Influence of a shift $\varepsilon $ of the centre of initial distribution of $r$
on (a) size $N_{min}$ of smaller of two subsets, which appear after reaching the Heider balance,
(b) the time $\tau$. As it is seen the plot (b), $\tau $ does not show any peculiar behaviour
near the transition point $\varepsilon \approx 0.036$, where the smaller subset vanishes.
These calculations are made for $N=100$.}
\label{tau_eps}
\end{center}
\end{figure}

All the results of Ref. \cite{my} and those reported above are obtained for the uniform initial
distribution of the matrix elements $r(i,j)$ with zero average. Within our sociological 
interpretation, this assumption can be interpreted as some symmetry of relations, with equal
number of their positive and negative values and intensities. This symmetry can be broken if
the initial values of $r(i,j)$ are randomly selected from the range 
$(-r_0+\varepsilon,r_0+\varepsilon)$. If $\varepsilon >0$, all the relations are improved;
if $\varepsilon <0$, all are deteriorated. If $\varepsilon $ is large enough, the output of 
the dynamics is that there is no division of the group. Instead, all relations become positive,
i.e. $r(i,j)>0$. A glance at Eq. 1 ensures that in this case all the matrix elements increase
in time; then this unity will continue forever. This is a kind of phase transition, with
$\varepsilon $ as a control parameter. The role of the order parameter can be assigned to the 
size $N_{min}$ of a smaller part of the group in the final state, when all triads are balanced. 
When the division does not take place, $N_min$ is zero. In Fig. 3 a, we show $N_{min}$ as 
dependent on $\varepsilon$. The plot is expected to be sharper if $N$ is larger, but 
the time of calculation increases with $N$ at least as $N^{3/2}$ and therefore the precision 
of determining the transition point is limited. As it can be seen in Fig. 3 a, this transition
point is close to $\varepsilon =0.036$. This means that accordance is reached when less than 4
percent of the matrix elements change their initial sign from negative to positive. 
We note that in the vicinity of the transition, 
$\tau $ does not show any discontinuity. This result is shown in Fig. 3 b. 

We checked that $\tau $
displays a maximum at $\varepsilon =0$. In particular, consider the case when 
$\varepsilon<-r_0$, i.e. all relations 
are initially hostile. Absolute values of some of them are large, and the time derivatives of 
some others are large as well. Then, the Heider balance is reached rather quickly. One could say
that a social state, when all relations are hostile, is rather unstable; soon we decide to  
collaborate with those who are less abominable, than the others.

\section{Discussion}

Main goal of this work is to develop the mathematical refinement \cite{my} of the problem of 
the Heider balance on networks. Already in previous works \cite{har,dor,wang}, the problem remained 
somewhat aside from the main stream of sociology, as taught in student textbooks \cite{tur,szt}. 
If the concept of the balance is taken as granted, the next step should be to include it into 
more established subjects, as the group dynamics \cite{col,kad} and the game theory \cite{str}.
However, such a task excesses the frames of this paper. 

It is tempting to interpret the results directly, taking literally all the associations
suggested above. There, the limitation $R$ can be seen as a variable which reflects the amount
of freedom allowed by the public opinion. If $R$ is large, strong feelings can be expressed freely 
and they influence emotions of the others. If $R$ is small, we speak mainly about the weather, and 
emotions remain hidden. By an appropriate shift of average initial social distance $\varepsilon$, we could 
manipulate the public opinion, at least temporarily preventing social instabilities. These analogies or metaphors can be continued,
but we should ask to what extent all of that can be realistic. This question is particularly
vivid when mathematical models are laid out for a humanistically oriented audience, when 
equations and plots bear some magic force of ultimate true. Doing the sociophysics, it is fair 
to make a short warning. Deeper discussion of relations between sociology and natural sciences 
can be found in Refs. \cite{mer,col}. 

Our experience with mathematical models is that their internal logic is false rarely, and errors
are easy to be detected and eliminated. Often, the problem is rather with the model assumptions and
interpretation - how they fit to reality? This point can be easily attacked by people
without mathematical education, and this kind of discussion can be equally instructive for both
sides. Indeed, a mathematical realization of a model is in many cases correct, but its 
assumptions and interpretation - input and output - are no more true when expressed with 
mathematical symbols, than with words. 

Bearing this in mind, we see the validity of our results mainly in improving the relation
between the Heider idea and its mathematical realization. We mean that in order to symbolize 
interpersonal
relations real numbers can be used with more sociological reality than integers. We mean
also that differential equations are more appropriate to describe the time dependence of 
human opinions, than just flipping units from positive to negative. We believe that
in the present version, the model is improved. Internal logic of its equations allows to 
draw the results, reported above. They may be true or not - it depends on how the model is used. 
We hope it can be useful in sociological applications.

\bigskip
{\bf Acknowledgements}. The authors are grateful to the Organizers and Participants
of the First Polish Symposium on Econo- and Sociophysics, where these results were reported
and discussed.

\bigskip

\end{document}